\documentstyle[aps,eqsecnum,preprint,multicol]{revtex}
\tightenlines

\begin{document}
\title{The Lennard-Jones--Devonshire cell model revisited}
\author{James E. Magee}
\address{Department of Physics and Astronomy, The University of Edinburgh,\\
Edinburgh EH9 3JZ, U.K.}

\author{Nigel B. Wilding} 
\address{Department of Mathematical Sciences, The University of
Liverpool, Liverpool L69 7ZL, U.K. }

\input epsf

%\tighten
\maketitle

\begin{abstract} 

We reanalyse the cell theory of Lennard-Jones and Devonshire and find
that in addition to the critical point originally reported for the
$12-6$ potential (and widely quoted in standard textbooks), the model
exhibits a further critical point. We show that the latter is actually
a more appropriate candidate for liquid-gas criticality than the
original critical point. 

\end{abstract}

%\draft


%\begin{multicols}{2}

\section{Introduction and background}

Lennard-Jones--Devonshire (LJD) cell theory  \cite{LJD,Kirkwood,Barker}
is a lattice approximation to the liquid state. Historically it was the
prototype microscopic model to predict the location of the
liquid-vapour critical point of a simple fluid and, as such, is widely
employed as an introduction to mean field theory in a number of
standard texts and papers\cite{Hill,FOWLER,BURSHTEIN}. It is still in
use today for research applications \cite{MODARRESS,S-L}.

\begin{figure}
\setlength{\epsfxsize}{12.0cm}
\centerline{\mbox{\epsffile{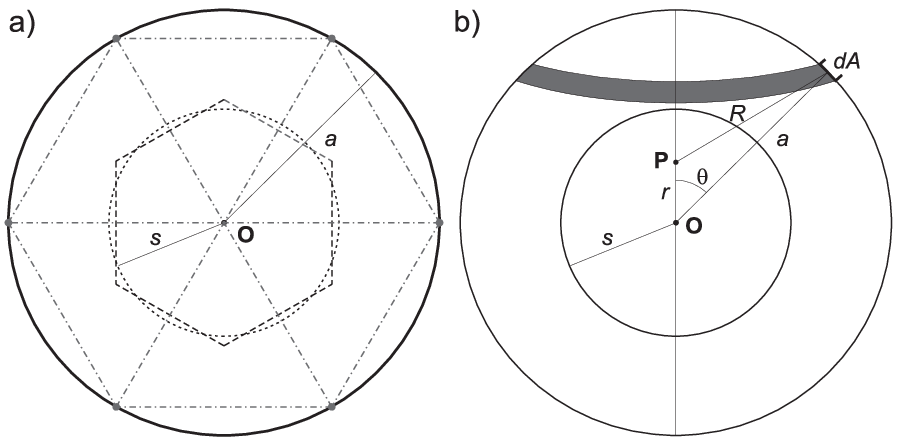}}} 
\vspace*{10mm}

\caption{ {\bf (a)} A cell (dotted circle) of volume $v$ and radius $s$
is centred on a lattice site with primitive unit cell (dashed hexagon)
of the same volume. The underlying lattice is shown with dot-dashed
lines, with lattice sites as vertices. The nearest neighbours are
assumed to be ``smeared out'' over the  interaction shell (see text),
of radius the lattice spacing $a$. {\bf (b)} The geometry for the
interaction of a particle at position $ \mathbf{P} $ within the cell
with an element of the shell $ dA $. Given the radial coordinate $ r $
of the particle, the angular coordinate $ \theta  $ of the element and
the shell radius $ a $, it is simple to calculate the separation $ R $
between particle and shell element; from this, the interaction can be
integrated around the shell.} \label{geomfig}

\end{figure}

Within the framework of the model, particles are considered to be
localised in singly occupied ``cells'', centred on the sites of a fully
occupied lattice (of some prescribed symmetry), within which they move
independently. For simplicity, cells are assumed to be spheres of
volume $v=V/N$ (the inverse number density) and radius $s$. A particle
is considered to interact with its $c$ nearest neighbours ``smeared''
around the surface of a further sphere of radius $a$ concentric with
the cell. The volume of this ``interaction sphere'' is related to the
cell volume by 

\begin{equation}
\label{gamma}
a^3=\gamma v\:,
\end{equation}
where $\gamma$ is a lattice-dependent constant, chosen so that for a
primitive unit cell of volume $v$ the lattice parameter will be the
radius of the interaction shell (see figure \ref{geomfig}(a)). 

The Helmholtz free energy of the model is given by \cite{LJD}

\begin{equation}
\label{cellthA}
A=-Nk_BT\ln v_{f}\sigma _{c}+\frac{NE_0}{2}\:.
\end{equation}
Here $E_0$ is the ``ground state energy'' - the energy per particle 
if all occupied their lattice sites; $\sigma_{c}$ is a constant
``communal entropy'' term\footnote{Reference \protect\cite{LJD} assigns
$\sigma_c=e$, although the exact value chosen for $\sigma _{c}$ has no
impact upon phase coexistence so long as it is independent of volume.},
which attempts to account for the entropy lost due to the localisation of particles
within cells; and $v_{f}$ is the ``free volume'',

\begin{equation}
\label{vf}
v_{f}=\int _ve^{-\left(E({\bf r})-E_0\right)/k_BT}d\mathbf{r}\:,
\end{equation}
with $ E(\mathbf{r})$ the ``cell potential'' the interaction
energy of a particle at a position $ \mathbf{r} $ within its cell.

Figure \ref{geomfig}(b) shows the cell geometry. The radial symmetry of
the system allows $E(\mathbf{r})$ to be written in terms of $E(r)$,
where $r$ is the radial coordinate of the particle's position within
the cell.  Simple trigonometry yields the separation $ R $ between the
particle at point $ \mathbf{P}$ within the cell and an element $d{\bf A}$
of the shell. If $c$ neighbouring particles are assumed to be smeared
over the interaction shell and the interparticle potential is given by
$u(R)$, the total energy of the particle $E(r)$ is given by

\begin{equation}
\label{E(r)}
E(r)=c\:\frac{\int _{shell}u(R)dA}{4\pi a^2}\:,
\end{equation}
where, for a prescribed cell size, the density enters through the
limits on $R$. One would normally perform this
integration numerically with further numerical integration to calculate
the free volume according to eq.~\ref{vf} for a given choice of number
density $\rho=v^{-1}$.

LJD used the model described above to calculate the phase diagram for the
$12$--$6$ Lennard-Jones (LJ) potential,

\begin{equation}
\label{12-6}
u(r)=4\varepsilon \left( \frac{\sigma ^{12}}{r^{12}}-\frac{\sigma ^{6}}{r^{6}}\right) \:.
\end{equation}
They assumed a face centred cubic lattice structure
with coordination number $c=12$ and geometrical constant $ \gamma =\sqrt{2}$.
The value of $E(r)$ (eq.~\ref{E(r)}) was calculated using the potential
of eq.~\ref{12-6}. However for the calculation of the ground state
energy $E_0$ (appearing in eq.~\ref{cellthA}) they used
the potential 

\begin{equation}
\label{12-6mod}
u(r)=4\varepsilon \left( \frac{\sigma ^{12}}{r^{12}}-1.2\frac{\sigma ^{6}}{r^{6}}\right) \:.
\end{equation}
Here the attractive part is increased by $20\%$, with respect to
eq.~\ref{12-6}, which was apparently motivated as
representing the effects of next and higher nearest neighbour interactions. 

For this system, LJD found a critical point at $k_BT_{c}/\varepsilon
\approx 1.3,\: P_{c}/\varepsilon\approx 0.6,\: \rho_c\approx 0.56 $. 
These results are to be compared with the consensus arising from a
number of simulation studies of criticality in the LJ fluid, the most
recent and sophisticated of which \cite{POTOFF} quotes $k_BT_{c}/\varepsilon=
1.3120(7)$, $P_{c}/\varepsilon=0.1279(6), \rho_c=0.316(1)$. That the
cell theory apparently predicts the critical temperature of the LJ
fluid to high accuracy has for many years been regarded as its major 
success \cite{Hill}. Indeed cell models remained the main method for
calculation of liquid equations of state up until the advent of density
functional theories \cite{BURSHTEIN}.

\section{Method and results}

We have obtained the cell theory phase diagram of the $12$-$6$ LJ
potential in order to compare with the results of LJD. Our
study employed the same model parameters as used in the original study
and was executed as follows. 

Using the Romberg method \cite{NRC}, numerical integration was carried
out to calculate the cell potential $E(r)$ and thence $v_f$ and
$H$ (cf. eqns~\ref{cellthA}--\ref{E(r)}). The equation of state in the
space of the pressure $P$, volume per particle $v$, and the temperature
$T$ was obtained by using a golden section search \cite{NRC} to
minimise the Gibbs free energy per particle 

\begin{equation}
\label{g}
g(v)=A/N+Pv\;,
\end{equation}
with respect to $v$ across a spread of points in $P,T$ space.  Within
this scheme, first order phase transitions were located by looking for
jumps in $v$ as $P$ and $T$ were varied\footnote{ Since the golden
section search is vulnerable to finding local minima in $g(v)$, we
supplemented this approach with a Newton-Raphson root-finding algorithm
\protect\cite{NRC}, which solves for the coexistence conditions
$g(v_{1})=g(v_{2})$, $P(v_1)=P(v_2)$.}. Critical points were estimated as the first
point along a phase boundary where no distinct coexistence volumes
could be found. 

\begin{figure}
\setlength{\epsfxsize}{12.0cm}
\centerline{\mbox{\epsffile{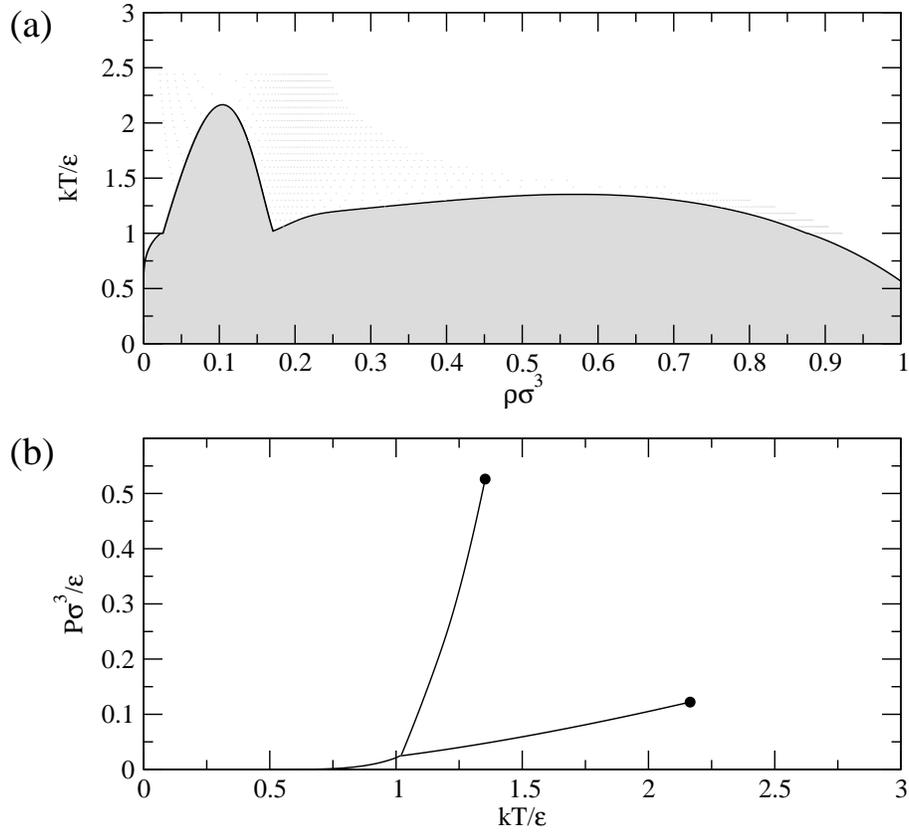}}} 

\vspace*{5mm}
\caption{ {\bf (a)} Projection of the cell
theory LJ potential phase diagram in the $\left( \rho ,T\right)$ plane
with $\rho=v^{-1}$. Dots show density points found using
the golden section search. Solid lines show coexistence densities.
{\bf (b)} $\left( P,T\right)$ projection of the same phase diagram. Solid lines
indicate lines of first-order phase transition, filled circles indicate
critical points. Note that the lower temperature, higher pressure
critical point is that found by LJD.}

\label{LJDphdiag}
\end{figure}

The results of implementing this procedure, are shown in figure
\ref{LJDphdiag}. We find a critical point at $k_BT_{c}/\varepsilon
=1.354, \:P_{c}/\epsilon=0.53,\: \rho_c=0.57$ which indeed appears to
duplicate the original LJD critical point \cite{LJD}. However, our full
phase diagram shows {\em two} lines of first order transition, meeting
at a triple point, \emph{both} terminating in critical points. 

In attempting to identify the phases appearing in figure
\ref{LJDphdiag}, it should be borne in mind that the lattice based
character of the cell model precludes an accurate representation of
non-crystalline phases. Accordingly it is not possible to assign
unambiguously the physical nature of the phases. Nevertheless a
plausible identification is {\em suggested} by the phase diagram itself
which, in the region of the triple point, is reminiscent of that for a
simple fluid such as Argon.  Above the triple point temperature in such
a system, an increase in pressure drives the system from a gas, to a
liquid to a solid. That the same phase assignments are reasonable in
the present model is supported by an analysis of their (radially
symmetric) cell potential $E(r)$, evaluated at the triple point
($k_BT_{c}/\varepsilon =1.02,P_{c}/\epsilon=0.025$),  which we now
address.

\begin{figure}
\setlength{\epsfxsize}{16.0cm}
\centerline{\mbox{\epsffile{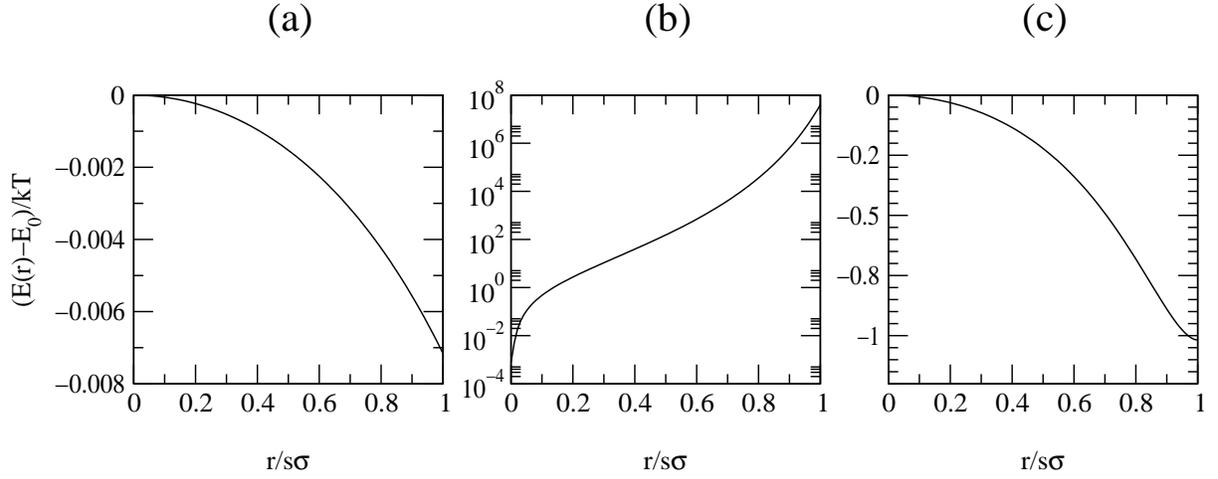}}}

\vspace*{5mm}

\caption{Radially symmetric cell potential $E(r)$ expressed relative to
the ground state energy $E_0$, for the modified cell theory with LJ
potential, calculated at the triple point from figure \ref{LJDphdiag}
{\bf (a)} Potential for the low density phase $( \rho /\sigma
^{3}=0.026 )$. {\bf (b)} Potential for the high density phase $( \rho
/\sigma ^{3}=0.869 )$ {\bf (c)} Potential for the intermediate density
phase $( \rho /\sigma^{3}=0.171)$, plotted from the centre of the cell,
$ r/s\sigma =0 $, to the edge, $ r/s\sigma =1 $, where $ s $ is the
radius of the cell at that density. }

\label{LJDpotns}
\end{figure}

Figure \ref{LJDpotns}(a) shows the cell potential $E(r)$ expressed
relative to the ground state energy $E_0$ for the low density phase.
This quantity is three orders of magnitude smaller than $k_BT$ across
the whole cell, so the low density phase can quite reasonably be
interpreted as gas-like. By contrast, the cell potential for the high
density phase (figure \ref{LJDpotns}(b)) exhibits a very steep minimum
at $r=0$, the centre of the cell. Away from the centre, the particles
experience the hard core repulsion of their neighbours and the
potential energy becomes large and positive. Thus the particle is
strongly confined to its lattice site, and the phase can be considered
solid-like. The cell potential energy for the intermediate density
phase is shown in figure \ref{LJDpotns}(c).  Its form is similar to
that of the gas-like phase though the scale is different--there is a
broad ``hump'' in the middle of the cell, of height around $k_BT$. Thus
the particle is still reasonably free within the cell and it seem
reasonable to interpret this as a liquid-like phase. With these phase
assignments \footnote{Hill \protect\cite{Hill} describes the cell
theory without the modified ground state term used by LJD, but quotes
the critical parameters corresponding to a modified system defined in
ref.~\protect\cite{WENTORF}. For the unmodified system we find cell
potentials similar in form to those of fig.~\ref{LJDpotns}, so we
retain our previous phase designations. However the critical
temperature of the ``solid-liquid''-like critical point found by LJD is
reduced to $k_BT_{c}/\varepsilon =0.82 $. The effect on the liquid-gas
transition is much less pronounced.}, the critical point found by LJD
is between a solid-like phase and a liquid-like phase, whilst the
other, lower-temperature critical point terminates a liquid-gas
transition at $k_BT_{c}/\varepsilon =2.17, P_{c}/\varepsilon=1.221,
\rho_c=0.104$. 

\section{Discussion and conclusions}

We have shown that the original LJD cell theory critical point for the
$12$-$6$ potential \cite{LJD} terminates a line of pseudo solid-liquid
coexistence, rather than a liquid-gas transition as originally
postulated.  Moreover, we find that the model exhibits a further first
order transition, which we believe is a more appropriate candidate for
the liquid-vapour line than that suggested by LJD.  It seems likely
that LJD overlooked the transition we have found because at the time of
their study, numerical integration was extremely labour intensive,
rendering it difficult to attain the rather low densities at which the
gas-like phase appears. Presumably also, the proximity of their
critical point temperature to the ``real'' LJ critical point (estimated
at the time from experimental data for Argon), made it tempting to make
the identification they did. In hindsight this agreement can only be
regarded as coincidental; one would expect a crude mean field theory
such as the cell model to considerably {\em overestimate} the critical
temperature due to the neglect of fluctuations. The new critical point
reported here does indeed do so.

Of course a critical point between a solid-like and liquid-like states
(as found by LJD) is an artifact of the cell structure of the model,
which imposes long ranged orientational and translational order on the
liquid where none exists in reality. The inability of cell theory to
accurately represent disordered phases has long been appreciated and
several attempts have been made to ameliorate this and other
shortcomings of the approach. These include zero- and
multiple-occupancy of cells (``hole'' theories) 
~\cite{Hill,HOLETHEORY}; calculations of interactions with second- and
higher nearest neighbours (see eg. ref.~\cite{WENTORF}); use of
numerical integration to give more accurate cell shapes and free
volumes ~\cite{BUEHLER}; and differing methods for calculating the cell
potential (see eg. \cite{MODARRESS}). However, none of these
extensions fully address the underlying lattice approximation which
renders suspect any cell theory equation of state for a liquid - by
definition, a disordered state without a lattice. Even hole theories
will only be able to represent a liquid of appreciable density as a
highly defective crystal.

Finally we briefly point out implications of our findings for
other recent work on cell theory. In ref.~\cite{MODARRESS} a modified
cell theory is introduced with the specific intention of engineering a
freezing transition. This authors state that no such transition exists
in the LJD cell theory, in the apparent belief that the critical point
of the bare model is liquid-gas in character. 

In other recent work (which in fact motivated the present study), a cell
model has been employed to study the phenomenon of the second critical
point which is observed in pure fluids interacting via so-called
core-softened potentials \cite{S-L}. In such systems liquid-liquid
phase separation occurs in addition to liquid-gas coexistence. Clearly
in view of our finding, that even the basic cell theory for the $12$-$6$
potential exhibits two critical points, the results of such studies
should be interpreted with caution. We intend to report in depth on
this matter in a future communication.

\acknowledgements

The authors thank A.D. Bruce, for helpful advice and discussions.

\newpage

\newpage

\end{document}